\RequirePackage{graphicx} 
\documentclass[final, num-refs]{wiley-article_final}
\usepackage[utf8]{inputenc}
\usepackage{amsmath}
\usepackage{bm}
\usepackage{bbold}
\usepackage{authblk}
\usepackage{amssymb}
\usepackage{graphicx}
\usepackage{ifdraft}
\usepackage{marginnote}
\usepackage{pdfpages}
\usepackage{ifpdf}
\ifpdf
	\usepackage[]{hyperref}
\else
	\usepackage[dvipdfm]{hyperref}
\fi
\usepackage{tikz}
\usepackage[
	capitalise,
]{cleveref}

\usepackage{etoolbox}
\usepackage{color}

\newtoggle{MRM}
\togglefalse{MRM}

\iftoggle{MRM}{
	\usepackage[nomarkers, nolists]{endfloat}
}

\usepackage{changes}
\newcommand{\stkout}[1]{\ifmmode\text{\sout{\ensuremath{#1}}}\else\sout{#1}\fi}
\setdeletedmarkup{\stkout{#1}}

\newcommand{\R}[1]{\ifoptionfinal{}{\marginnote{\textcolor{blue}{#1}}}}
\renewcommand{\b}{\ifoptionfinal{}{\textcolor{blue}}}

\usepackage[
per-mode=fraction,
separate-uncertainty,
retain-unity-mantissa=false,
product-units=power,
table-alignment=center,
detect-all,
binary-units=true,
exponent-product=\cdot,
exponent-base=10
]{siunitx}

\papertype{Note}
\title{{Simple Auto-Calibrated Gradient Delay Estimation From Few Spokes 
Using Radial
Intersections (RING)}}

\author[1\authfn{1}]{Sebastian Rosenzweig}
\author[1,2]{H. Christian M. Holme}
\author[1,2]{Martin Uecker}
\affil[1]{Institute for Diagnostic and Interventional Radiology,
	University Medical Center Göttingen, Göttingen, Germany}
\affil[2]{German Centre for Cardiovascular Research (DZHK),
	Partner site Göttingen, Göttingen, Germany}

\corraddress{Sebastian Rosenzweig, University Medical Center Göttingen,
	Institute for Diagnostic and Interventional Radiology,
	Robert-Koch-Str. 40,
	37075 Göttingen, Germany }
\corremail{sebastian.rosenzweig@med.uni-goettingen.de\\
	
	Submitted to Magnetic Resonance 
	in 
	Medicine as a Note.
	\bf{{Word count}}: Abstract 170, Body 2921.}

\begin{document}
\runningauthor{Rosenzweig et al.}

\maketitle

\begin{abstract}
{\bf Purpose:} 
To develop a simple and robust tool for the estimation of gradient delays
from highly undersampled radial k-space data. \\
{\bf Theory:}
In radial imaging gradient delays induce parallel and orthogonal trajectory 
shifts, which can be described using an ellipse model. The intersection points 
of the radial spokes, which can be estimated by spoke-by-spoke comparison of
k-space samples, distinctly determine the parameters of the ellipse.
Using the proposed method (RING), these parameters can be obtained using a
least-squares fit and utilized for the correction of gradient delays. \\
{\bf Methods:}
The functionality and accuracy of the proposed RING method is validated and 
compared to correlation-based gradient-delay estimation from opposing
spokes using numerical simulations, phantom and in vivo heart measurements.\\
{\bf Results:} 
In all experiments, RING robustly provides accurate gradient delay 
estimations even for as few as three radial spokes.\\
{\bf Conclusion:}
The simple and straightforward to implement RING method provides accurate 
gradient delay estimation for highly undersampled radial imaging.
\end{abstract}

\section{Introduction}

In todays clinical practice almost all magnetic resonance imaging (MRI)
techniques are based on Cartesian trajectories. Nevertheless, in the recent 
years
non-Cartesian trajectories, in particular radial acquisitions, drew
increasing interest among the scientific community for
their motion robustness and milder undersampling artifacts and proofed
potential for a significant imaging speed-up~\cite{Song_Magn.Reson.Med._2000,Mistretta_Magn.Reson.Med._2006,Uecker_Magn.Reson.Med._2010,Wright_J.Magn.Reson.Imaging_2014,Block_J.KoreanSoc.Magn.Reson.Med._2014}.
Although the benefits of radial trajectories seem evident, for various reasons
they did not yet find widespread application in clinical routine. One reason
is the higher sensitivity to system imperfections such as eddy-current-induced
gradient delays, that lead to deviations from the nominal
sampling locations~\cite{Peters_Magn.Reson.Med._2003}. To tackle this problem,
a variety  of different trajectory error compensation strategies have been
developed.\\
One  approach is to measure the actual k-space trajectory during each measurement
using specialized hardware, which is highly effective but
expensive and not always practical \cite{Barmet_Magn.Reson.Med._2008,Dietrich_Magn.Reson.Med._2016}.
Alternatively,  calibration scans can be
utilized to fully characterize the gradient system impulse response function
(GIRF) of the scanner~\cite{Vannesjo_Magn.Reson.Med._2013,Liu_Materials_2013,Jang_Magn.Reson.Med._2016,Stich_Magn.Reson.Med._2018}.
However, as these methods have not yet been adopted by the vendors they
require significant implementation efforts and are difficult to
integrate into a complete workflow. Furthermore, GIRF-based methods
cannot capture sequence or protocol dependent temporal variations of
the gradient system e.g. through heating, and the characterization must
be repeated regularly to compensate long-term system variances.\\
Other approaches have been designed
particularly for error compensation in radial imaging. Iterative, parallel
imaging-based methods developed by Deshmane et al.\
\cite{Deshmane_Magn.Reson.Med._2016} and Wech et al.\
\cite{Wech_MagnResonMed_2015} exploit correlations in the receive channels
using GRAPPA operator gridding \cite{Seiberlich_Magn.Reson.Med._2007} and
shift the samples or the trajectory position until a certain condition
is fulfilled. Both methods are data driven and therefore can be used to
retrospectively detect and compensate transient trajectory errors.
These methods require good estimates for
the gridding operators, which in turn demands sufficient in-plane coil
sensitivity variation and a certain number of spokes for auto-calibration. 
These conditions might not always be given, as e.g. in
interactive real-time MRI
the imaging plane is repeatedly being rotated and shifted 
\cite{Kerr_Magn.Reson.Med._1997,Guttman_Magn.Reson.Med._2003, 
Yang_JAmCollCardiol_1998,Unterberg-Buchwald_J.Cardiov.Magn.Reson._2017}.
Recently, Jiang et al.\ \cite{Jiang_Magn.Reson.Med._2018} introduced an
framework which simultaneously estimates gradient delays and coil
sensitivities using an alternating minimization approach. This method inspired
by SAKE \cite{Shin_Magn.Reson.Med._2014} uses a computationally rather
demanding low-rank constraint in conjunction with the Gauss-Newton method to
solve a non-linear optimization problem. Furthermore, two more general
algorithms have been proposed that combine
trajectory correction and image reconstruction~\cite{Mani_Magn.Reson.Med._2018,Ianni_Magn.Reson.Med._2016}.
 However, in the
approach of Mani et al.\ \cite{Mani_Magn.Reson.Med._2018} prior knowledge about
the coil sensitivities is necessary. 
Moreover, both algorithms might be difficult to combine with
other image reconstruction methods.\\
A noticeable alternative to the above mentioned rather elaborate trajectory
correction techniques is the {Adaptive method} developed by Block and
Uecker \cite{Block__2011} for
radial imaging. Because of its intuitive
approach, its robustness and accuracy and the straight forward
implementation it found widespread acceptance and application 
\cite{Feng_Magn.Reson.Med._2014,Wundrak_IEEETransMedImag_2015, 
Moussavi_Magn.Reson.Med._2013,Untenberger_Magn.Reson.Med._2016,Block_J.KoreanSoc.Magn.Reson.Med._2014}.
The method requires calibration scans where pairs of opposed spokes
with varying orientation are acquired. Then, the sample shift along the readout
direction is calculated by
performing a cross-correlation of the opposed spokes. The shifts are fitted to
a linear ellipse model
\cite{Peters_Magn.Reson.Med._2003,Moussavi_Magn.Reson.Med._2013} and
compensated for in the gridding procedure.
Recently, it was shown that the same approach can be used to estimate the
gradient delays from the data itself without the need for calibration scans. 
\b{This auto-calibrated variant, here dubbed AC-Adaptive method, was described 
by Untenberger et al.\ \cite{Untenberger_Magn.Reson.Med._2016}, applied in 
several publications, e.g. 
\cite{Wang_Magn.Reson.Med._2018,Schaetz_ComputMathMethodM_2017,Volkert_Int.J.Imag.Syst.Tech._2016},
 and 
studied 
in \cite{Rosenzweig__2018}. Instead of using perfectly anti-parallel spokes, 
the AC-Adaptive method takes spokes from the actual radial acquisition which 
are only approximately anti-parallel to estimate the shifts.} \R{Ref 1.1: We 
clarified the idea of the AC-Adaptive method and cite papers using it.} This 
allows for 
real-time and retrospective gradient delay
correction which can be applied to e.g. interactive real-time
MRI or to compensate gradient delay changes due to coil heating.  However, the
present work shows that the AC-Adaptive method is not fully consistent with the
ellipse model which it is based on. In settings with oblique slices and/or non-isotropic
delays of the physical gradients, the spokes will experience both a shift in
read-out direction and orthogonal to it. Due to the latter, even perfectly
opposed spokes will not cover the same k-space line which leads to
inconsistencies in the cross-correlation calculation. Furthermore, the
AC-Adaptive method cannot provide stable gradient delay
estimates given only very few spokes, as then no nearly opposed spoke pairs
exist. \\
The aim of this work is to develop a simple yet accurate and robust gradient delay
estimation tool which also
works for very few spokes. We show that our method, which uses Radial spoke
INtersections for Gradient delay estimation (RING), outperforms the AC-Adaptive
method in all investigated numerical simulations, phantom and in vivo heart
measurements and provides precise gradient delay estimates even for as few as
$3$ spokes.

\section{Theory}
\begin{figure}[h]
	\centering
	\includegraphics[width=0.6\textwidth]{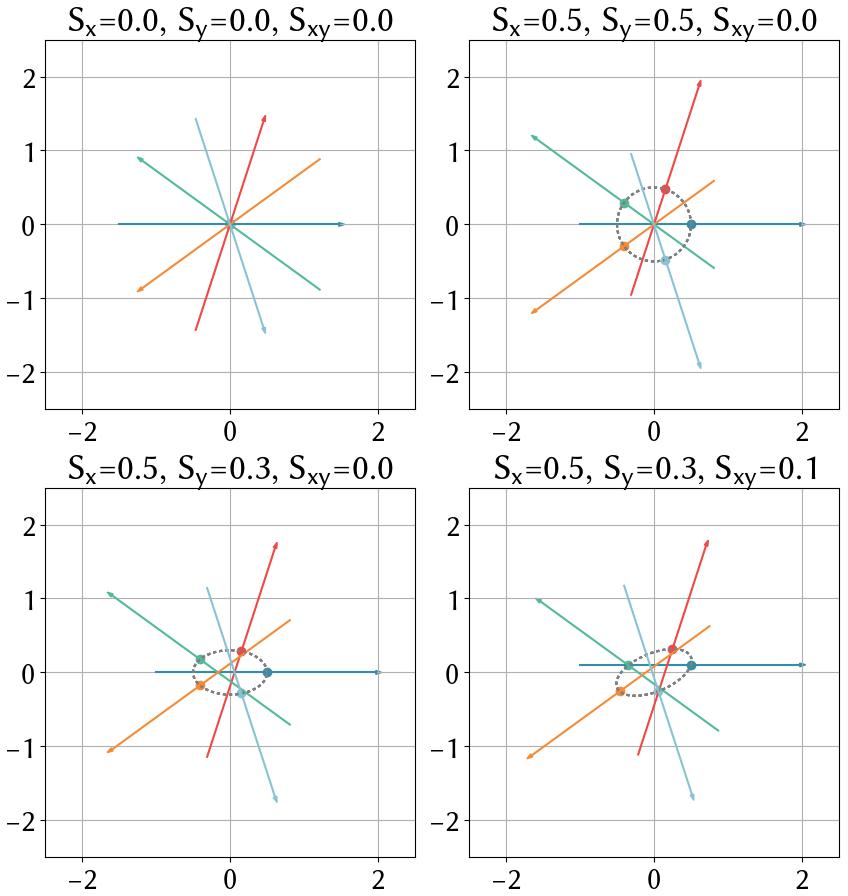}
	\caption{Schematic of shifted k-space trajectories in units of 
	$1/\text{FOV}$ 
	for different delays $S_\text{x}$, $S_\text{y}$ and $S_\text{xy}$. 
	Exemplarily, 5 abridged spokes are depicted. The 
	arrows point in readout direction and the DC sample of each spoke is 
	highlighted by a dot. The ellipse determined by $\delta \bm{k}$ (Eq.\ 
	(\ref{Eq:deltak})) is plotted as a dashed gray circle.}
	\label{Fig:Trajplot}
\end{figure}

\paragraph{The gradient delay ellipse model.}
Peters et al.\ showed that in radial imaging, linear eddy current effects delay
the start of the 
readout gradients which induces both a parallel and an orthogonal shift to the 
nominal k-space trajectory, while the projection direction is not 
affected~\cite{Peters_Magn.Reson.Med._2003}. Moussavi et al.\ 
\cite{Moussavi_Magn.Reson.Med._2013}
proposed a simple model to describe 
this spoke shift as a vector
\begin{equation}
\delta \bm{k} := {\bm{S}} \hat{\bm{n}}_\theta,
\label{Eq:deltak}
\end{equation}
\begin{equation}
\hat{\bm{n}}_\theta:=\left( 
\begin{array}{c} 
\cos\theta \\ \sin\theta\end{array} \right),
\label{Eq:ProjDir}
\end{equation}
\begin{equation}
\bm{S} := \left( 
\begin{array}{cc}
S_x & S_{xy} \\
S_{xy} & S_y 
\end{array}
\right)
\label{Eq:S}
\end{equation}
with $\theta$ the projection angle and $\hat{\bm{n}}$ the normalized projection 
direction. $S_x$ and $S_y$ capture the delays in the axial case whereas 
$S_{xy}$ guarantees three-dimensional rotational invariance and particularly 
accounts for the interaction of all three physical gradients when measuring 
oblique slices. For a detailed derivation please consider the appendix. The 
goal of this work is to efficiently determine the 
parameters $S_\text{x}$, $S_\text{y}$ and $S_\text{xy}$, which can then be used 
to calculate the actual (shifted) trajectory needed for accurate gridding in 
image reconstruction.\\
Let $N_\text{samp}$ be the number of samples in readout direction and  
$N_\text{sp}$ the number of spokes used for gradient delay estimation. Then, 
the 
sample positions in units of $1/\text{FOV}$ of a spoke with projection angle 
$\theta_i$, $i\in\{1,\dots,N_\text{sp}\}$, can be modeled 
using the parametric linear equation
\begin{equation}
\bm{r}_{\theta_i}= {\bm{S}}\hat{\bm{n}}_{\theta_i} + 
a_{\theta_i} \hat{\bm{n}}_{\theta_i}
\end{equation}
with $a_{\theta_i}=[-N_\text{samp}/2,N_\text{samp}/2-1]$. 
The ellipse defined by Eq.\ (\ref{Eq:deltak}) determines the 
position $\bm{r}_{\theta_i}(a_{\theta_i}=0)$ of the shifted spokes.\\
Fig. \ref{Fig:Trajplot} 
depicts a schematic of actual k-space trajectories for different delays
${\bm{S}}$. If no gradient delays are present (top-left) the spokes are 
not shifted at all. For isotropic delays in the axial case (top-right) the 
spokes are translated in readout direction only. For anisotropic delays 
(bottom-left) and/or oblique slices (bottom-right) the spokes experience both a 
readout-shift and an orthogonal shift and no longer intersect in the k-space 
center. The intersection points of the spokes relative to their DC 
component
uniquely define the shift matrix ${\bm{S}}$.
\paragraph{Determination of $S_x,\;S_y\;\text{and}\; S_{xy}$.}
The intersection point of the spokes $\bm{r}_{\theta_i}$ and 
$\bm{r}_{\theta_j}$ 
yield a conditional equation for $\bm{S}$:
\begin{gather}
\bm{r}_{\theta_i}(a'_{\theta_i}) \overset{!}{=} 
\bm{r}_{\theta_j}(a'_{\theta_j}), \\
\bm{S}(\hat{\bm{n}}_{\theta_i} - \hat{\bm{n}}_{\theta_j}) = 
a'_{\theta_j}\hat{\bm{n}}_{\theta_j} - 
a'_{\theta_i}\hat{\bm{n}}_{\theta_i}.
\label{Eq:Sgeneral}
\end{gather}
 To facilitate calculations, we introduce the definitions 
\begin{equation}
\bm{s}:=\left( \begin{array}{c}
S_x\\
S_y\\
S_{xy}
\end{array}\right),
\end{equation}
\begin{equation}
\hat{\bm{n}}_{\theta_i} - \hat{\bm{n}}_{\theta_j} := 
\left(
\begin{array}{c}
\xi_1\\
\xi_2
\end{array}\right),\;\; \bm{A} := \left(\begin{array}{ccc} 
\xi_1 & 0 & \xi_2\\
0 & \xi_2 & \xi_1
\end{array}
\right),
\label{Eq:Equality}
\end{equation}
\begin{equation}
\bm{b} := a'_{\theta_j}\hat{\bm{n}}_{\theta_j} - 
a'_{\theta_i}\hat{\bm{n}}_{\theta_i},
\end{equation}
and rearrange Eq.\ (\ref{Eq:Sgeneral})
\begin{equation}
\bm{A}\bm{s}=\bm{b}.
\label{Eq:InverseProblem}
\end{equation}
If $N_\text{sp}>2$, Matrix $\bm{A}$ and vector $\bm{b}$ can be extended to 
contain all 
considered intersection points. Then, the system of equations Eq.\ 
(\ref{Eq:InverseProblem}) for
$\bm{s}$ is overdetermined and $\bm{s}$ can be 
obtained by a least-squares fit using the pseudo-inverse
\begin{equation}
\bm{s} = (\bm{A}^T\bm{A})^{-1} \bm{A}^T \bm{b},
\label{Eq:s}
\end{equation}
where $^T$ indicates the transpose.
\paragraph{Determination of intersection points.}
To obtain $\bm{s}$ (Eq.\ (\ref{Eq:s})) the values for 
$a'_\theta$ in Eq.\ 
(\ref{Eq:Sgeneral}) must be determined from the measured data. Since the 
k-space 
value at the intersection point of two spokes should be identical except for 
noise in all channels, the values for $a'_\theta$ can be obtained by 
comparing the actual complex
sample values of the spokes. The sample pair for which the root-sum-of-squares 
difference over all channels is minimal is assumed to represent the 
intersection point. To guarantee accurate estimates for 
$a'_\theta$ and thus $\bm{s}$, each spoke is retrospectively 
sub-sampled \b{via Fourier interpolation} and denoised: \R{Ref 1.5: We have 
added the keyword 'Fourier 
	interpolation' to mention the technique that we use for sub-sampling. The 
	exact 
	procedure is described after the semicolon.} \b{An inverse Fourier 
transform is used to obtain a spoke's image 
domain representation.} As the readout direction is generally oversampled by a 
factor of 2, all samples outside of the central 
$N_\text{samp}\times 0.6$ region can be set to zero to sinc-denoise k-space. 
\b{Then, the 
data is zero-padded by 
$N_\text{pad} \times N_\text{samp}$ (we propose $N_\text{pad}=100$) and the 
sub-sampled k-space 
is retrieved using another 
Fourier transform.} As most of the 
energy is localized in the low spatial frequency region it is sensible to 
investigate only intersection points in the central region of k-space, which 
avoids
inaccuracies due to noise. Therefore, only the intersection point of a
spoke with its most orthogonal counterpart is considered, \b{i.e. we search for 
the spoke pairs that best approximate an intersection angle of 
$\SI{90}{\degree}$.} Then, the
sample-wise comparison can be restricted to the central \b{$\beta\times 
N_\text{pad}$}\R{Ref 1.6: Clarification of the term 'most orthogonal 
counterpart'. \\
Ref 1.7: We changed the symbol $\gamma$ to $\beta$.\\
Ref 1.4: See supplementary material}
(we 
propose 
$\beta=1.5$) samples of the spokes.

\b{An analysis of the accuracy of the proposed method to determine the 
intersection points for different noise values and simulated phantoms is 
provided as supplementary material.}

\section{Methods}
All measurements were performed on a SIEMENS Skyra 3T scanner, all 
reconstructions and the gradient delay estimation methods were implemented and 
performed using BART \cite{Uecker__2015}. In our study we chose the RING 
parameters to be $N_\text{pad}=100$ and $\beta=1.5$, which provided accurate 
and robust results throughout all experiments. In  the  interest  of  
reproducible  
research,  code  and  data  to reproduce the experiments are made available on 
Github.\footnote{\url{https://github.com/mrirecon/RING}}\R{Ref 1.5: All code 
and data will be made publicly available after successful publication.}

\paragraph{Numerical Simulations.}
To demonstrate the general functionality and accuracy of our method we 
performed a numerical Shepp-Logan k-space phantom study (oversampled readout 
samples 128, 8 channels) using a golden angle scheme and spokes shifted 
according to Eq.\ 
(\ref{Eq:deltak}) 
with the nominal values $\bm{s}^\text{iso}=(0.3,0.3,0)$, 
$\bm{s}^\text{ax}=(0.3,-0.1,0)$ and 
$\bm{s}^\text{obl}=(0.3,-0.1,0.2)$ respectively.\\
To quantify the estimation 
error $\mathcal{E}(\bm{s}, \bm{s}^\text{est}_{N_\text{sp}})$ we used the L2 
norm 
\begin{equation}
\mathcal{E}(\bm{s},\bm{s}_{N_\text{sp}}^\text{est}) := \sqrt{(S_x - 
S_x^\text{est})^2 + 
(S_y - 
S_y^\text{est})^2 + 
(S_{xy} - S_{xy}^\text{est})^2},
\label{Eq:Error}
\end{equation}
where $S$ stands for the nominal and $S^\text{est}$ for the estimated shifts. 
We estimated the shifts using RING for all numbers of spokes in the range 
$N_\text{sp}\in [3,127]$ 
and 
performed the simulations for projection angles distributed over a 
half circle ($\theta_i \in [0,\pi]$) and a full circle ($\theta_i\in[0,2\pi]$) 
\cite{Haji-Valizadeh_Magn.Reson.Med._2018}.\\
For comparison we did the same experiments using the AC-Adaptive method.

\paragraph{Phantom measurements.}
For the measurement on a custom-made brick phantom we used the SIEMENS 
Head-Neck-20 coil and a FLASH sequence (FOV = $256 \times 
\SI{256}{\square\milli\meter}$, oversampled readout samples = $320$, number of 
spokes 
= $159$, TE/TR = 
$1.46/\SI{2.3}{\milli\second}$, slice thickness = $\SI{5}{\milli\meter}$) with 
golden angle (half and full circle) acquisition. For the sake 
of better visibility of gradient delay artifacts, only $39$ spokes were used 
for image reconstruction. The delays were estimated for all numbers of 
spokes in the range $N_\text{sp}\in [3,159]$ using both RING  and
the 
AC-Adaptive method for full circle acquisitions. For half circle acquisitions 
only 
RING is utilized. We used radial 
NLINV \cite{Uecker_Magn.Reson.Med._2008,Uecker_Magn.Reson.Med._2010}
in combination with the corrected trajectories for image 
reconstruction.\\
For each number of spokes $N_\text{sp}$ the estimated delays 
$\bm{s}^\text{est}_{N_\text{sp}}$ were compared to $\bm{s}^\text{est}_{159}$ 
and the L2 errors 
$\mathcal{E}(\bm{s}^\text{est}_{159},\bm{s}^\text{est}_{N_\text{sp}})$ 
according to 
Eq.\ 
(\ref{Eq:Error}) were calculated.

\paragraph{In vivo measurements.}
We performed an in vivo measurement on a human heart (short-axis view, 30 
channel thorax and spine coil, FLASH sequence, FOV = $256 \times 
\SI{256}{\square\milli\meter}$, oversampled readout samples = $320$, TE/TR = 
$1.47/\SI{2.3}{\milli\second}$, slice thickness = $\SI{8}{\milli\meter}$) using 
a full circle golden angle acquisition scheme. $75$ consecutive spokes during 
the end-diastole were combined for image reconstruction with ENLIVE 
\cite{Holme_ArXiv_2017}. The 
gradient delays were estimated using RING and the AC-Adaptive method 
utilizing all numbers of spokes in the range $N_\text{sp}\in[3,75]$. The L2 
errors were calculated as described previously.\\
Human imaging was approved by the local ethics committee. Written informed
consent was obtained from the subject before the imaging.

\section{Results}
\paragraph{Numerical simulations.}

\begin{figure}[h]
	\centering
	\includegraphics[width=1\textwidth]{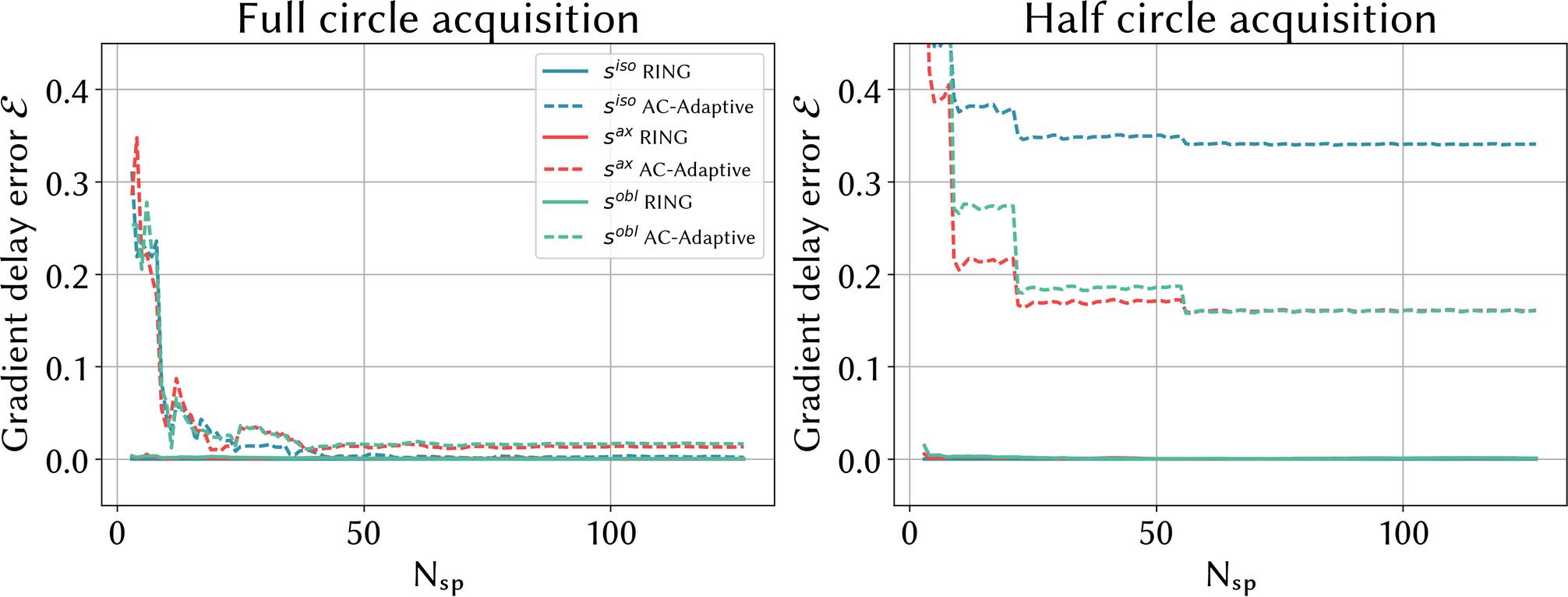}
	\caption{Gradient delay error 
	$\mathcal{E}(\bm{s},\bm{s}^\text{est}_{N_\text{sp}})$  (Eq. 
	(\ref{Eq:Error}))
		against the number of spokes 
		utilized for gradient delay estimation using the RING and the 
		AC-Adaptive 
		method. Numerical k-space 
		Shepp-Logan-phantom study with different nominal k-space trajectory 
		shifts 
		$\bm{s}^\text{iso}=(0.3,0.3,0)$, $\bm{s}^\text{ax}=(0.3,-0.1,0)$ and 
		$\bm{s}^\text{obl}=(0.3,-0.1,0.2)$. a) Full circle golden angle 
		acquisition (projection angle $\theta_i\in[0,2\pi]$). b) Half circle 
		golden angle acquisition (projection angle $\theta_i\in[0,\pi]$).}
	\label{Fig:Sim}
\end{figure}

The results of the numerical simulations are depicted in Fig. \ref{Fig:Sim}. 
The error $\mathcal{E}(\bm{s},\bm{s}^\text{est}_{N_\text{sp}})$ of the 
estimated 
gradient 
delays for different trajectory shifts using the AC-Adaptive method and RING 
are plotted over the number of spokes used for gradient delay 
estimation. \\
The RING method provides nearly perfect estimates for all 
investigated delays in half circle and full circle acquisitions, even if only 
three spokes are employed.\\
In contrast, the AC-Adaptive method delivers inaccurate gradient delay estimates for 
half circle acquisitions even if up to $N_\text{sp}=127$ spokes are used.
For full circle 
acquisitions at least $N_\text{sp}\approx20$ spokes are necessary, to provide 
reasonable 
results. For fewer numbers of spokes the gradient delay error shows 
unpredictable behavior, which makes the estimates unreliable. By increasing 
the number of utilized spokes, the estimated delays converge to constant 
values. However, only the isotropic delay is estimated perfectly, whereas in 
the axial and oblique case a deviation from the optimal values remains.

\paragraph{Phantom measurements.}

\begin{figure}[h]
	\centering
	\includegraphics[width=1\textwidth]{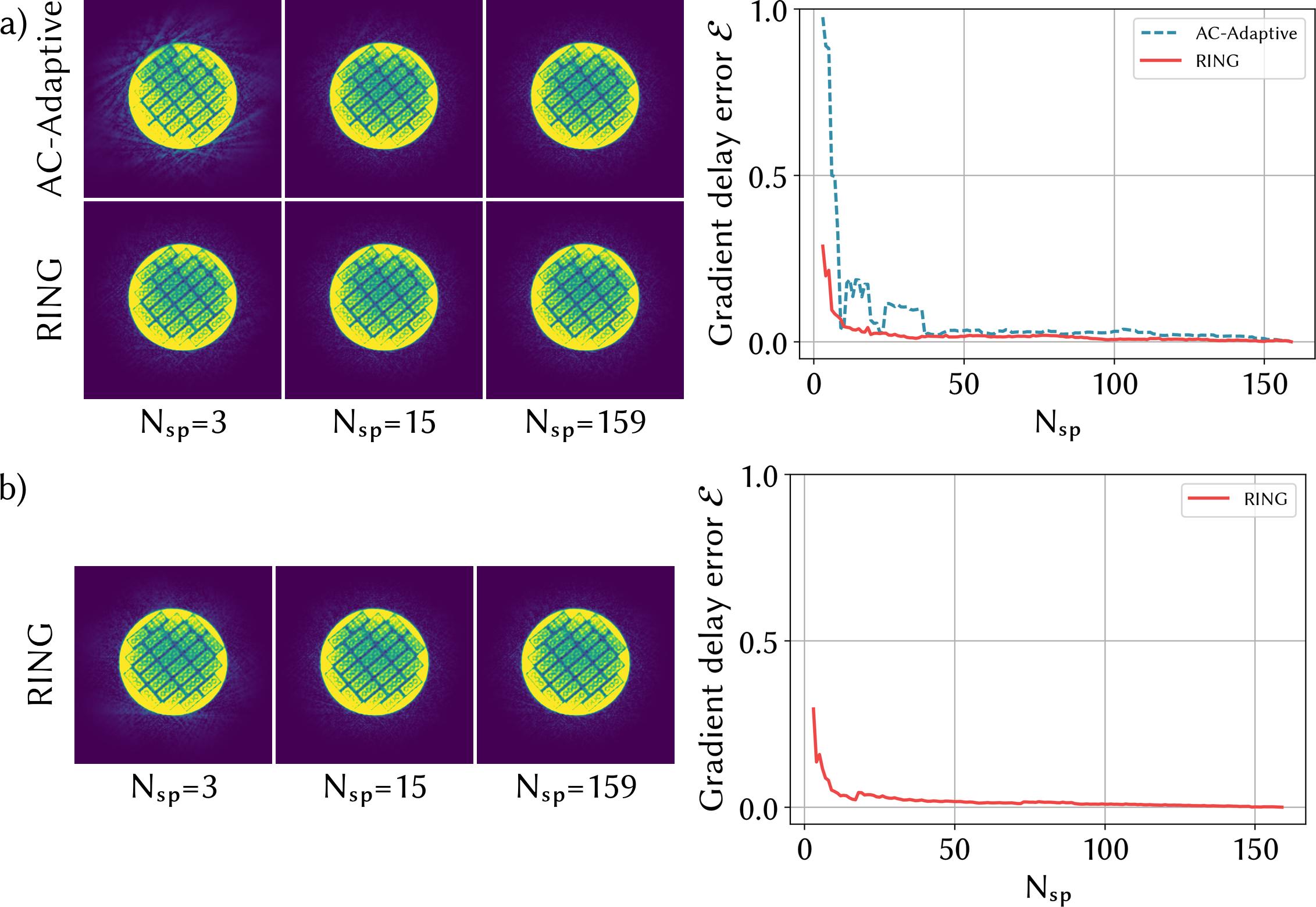}
	\caption{NLINV reconstructions using 39 spokes of a brick phantom 
		FLASH measurement with gradient delay correction estimated from 
		$N_\text{sp}$ 
		number of 
		spokes utilizing the AC-Adaptive method and the RING method. Besides, 
		the 
		gradient 
		delay error 
		$\mathcal{E}(\bm{s}^\text{est}_{159},\bm{s}^\text{est}_{N_\text{sp}})$ 
		(Eq.\ 
		(\ref{Eq:Error})) against the number of spokes used for gradient delay 
		estimation is depicted. a) Full circle 
		golden 
		angle acquisition (projection angle $\theta_i\in[0,2\pi]$). b) Half 
		circle 
		golden angle acquisition (projection angle $\theta_i\in[0,\pi]$). }
	\label{Fig:Lego}
\end{figure}

The results of the phantom measurement and the corresponding gradient delay 
errors 
$\mathcal{E}(\bm{s}^\text{est}_{159}, \bm{s}^\text{est}_{N_\text{sp}})$ over 
the number of 
spokes 
used 
for gradient delay estimation are provided in Fig. \ref{Fig:Lego}.\\
RING provides a good gradient delay estimation and thus, effective 
streaking artifact reduction even for only $N_\text{sp}=3$ spokes. The estimate 
is further improved when more spokes are utilized. However, the effect on the 
resulting image quality is only marginal, since the image is 
basically streaking free for $N_\text{sp}=3$ spokes already. Again, the method 
proves to 
be applicable to both full circle and half circle acquisition, although visual 
observation reveals slightly better results for full circle acquisitions 
with very few utilized spokes.\\
The AC-Adaptive method can only be applied in the full circle case, but does not
provide accurate gradient delay estimates for few numbers of spokes where the 
results appear worse than the uncorrected image (not shown). Although the 
actual convergence value is not reached until $N_\text{sp}=37$ spokes, the 
results for $N_\text{sp}=15$ already look 
suitable, even if some streaking artifacts can still be observed at the 
top.\\
The actual gradient delay is not know but the convergence values for full 
circle acquisitions,
$\bm{s}^\text{AC-Adaptive}_{159}=(0.336,0.360,-0.021)$ and
$\bm{s}^\text{RING}_{159}=(0.345,0.384,-0.009)$, are very similar for both 
methods. \b{NUFFT reconstructions show an equivalent behavior and are depicted 
in  
Supplementary Figure 2.} \R{Ref 2}

\paragraph{In vivo measurements.}

\begin{figure}[h]
	\centering
	\includegraphics[width=1\textwidth]{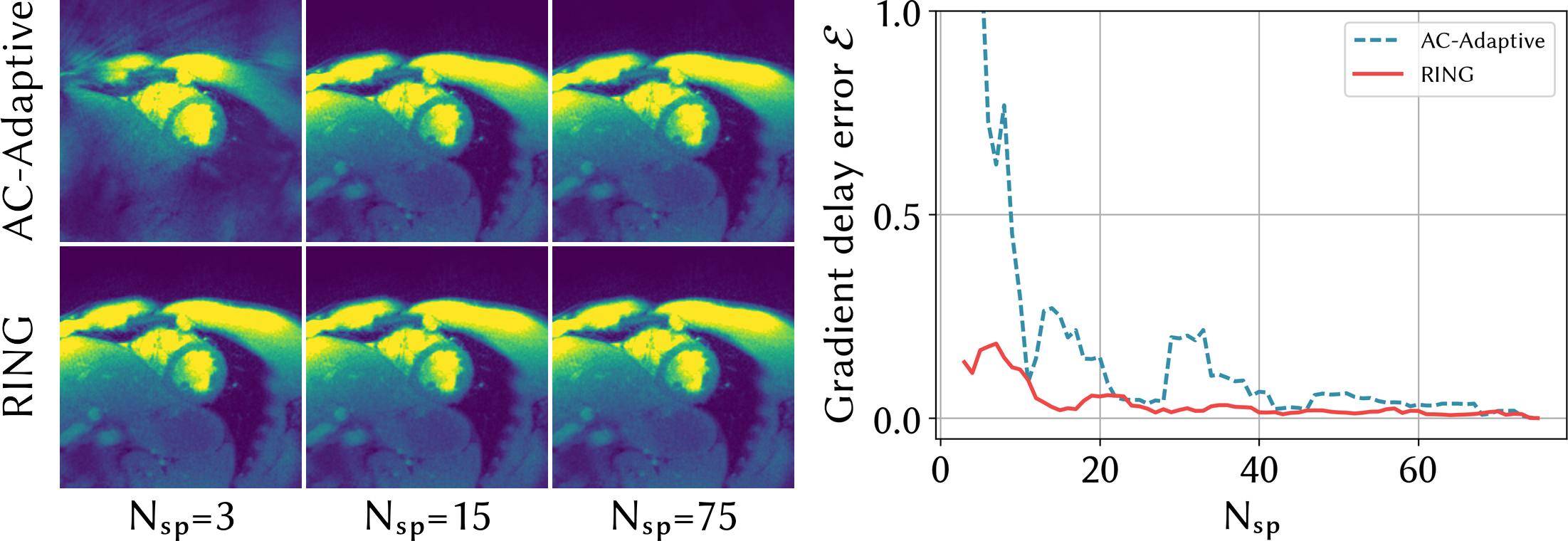}
	\caption{ENLIVE reconstructions using 75 spokes of an in vivo FLASH 
		measurement of the human heart (short-axis view, end-diastole)
		with gradient delay correction estimated from $N_\text{sp}$ 
		number of 
		spokes utilizing the AC-Adaptive method and the RING method. Besides, 
		the 
		gradient 
		delay error 
		$\mathcal{E}(\bm{s}^\text{est}_{75},\bm{s}^\text{est}_{N_\text{sp}})$ 
		(Eq.\ 
		(\ref{Eq:Error})) against the number of spokes used for gradient delay 
		estimation is depicted.}
	\label{Fig:Invivo}
\end{figure}

Figure \ref{Fig:Invivo} shows the effects of the gradient delay correction on 
an in vivo measurement of the human heart, as well as the gradient delay error 
$\mathcal{E}(\bm{s}^\text{est}_{75},\bm{s}^\text{est}_{N_\text{sp}})$ for both 
the AC-Adaptive method and RING.\\
Using RING, even $N_\text{sp}=3$ are sufficient for effective 
streaking artifact elimination and the gradient delay error compared to the 
convergence value is small for all investigated numbers of spokes.\\
The AC-Adaptive method also provides robust streaking suppression for 
$N_\text{sp}>40$. It estimates gradient delays close to convergence value 
for some smaller numbers of spokes, however, these 
results are not reliable as the gradient delay error shows large  
oscillations. For very few numbers of spokes, e.g. $N_\text{sp}=3$, the 
estimates are useless and even amplify streaking artifacts. \\
The convergence values of the gradient delay estimates are 
$\bm{s}_{75}^\text{RING}=(0.178,0.068,0.068)$ and 
$\bm{s}_{75}^\text{AC-Adaptive}=(0.263,-0.032,0.080)$ and yield comparable 
image quality.
\b{NUFFT reconstructions show an equivalent behavior and are depicted in 
Supplementary Figure 3.} \R{Ref 2}

\section{Discussion}
In this work we compared the widely-used AC-Adaptive method without 
calibration scans 
\cite{Block__2011,Rosenzweig__2018,Untenberger_Magn.Reson.Med._2016}
for 
gradient delay estimation in radial imaging with the here introduced Radial 
INtersection Gradient delay estimation (RING) method. The advantage of these 
two methods compared to other trajectory correction approaches is
the simple and straightforward implementation while still being robust and 
accurate. 
All investigated experiments revealed that RING outperforms 
the AC-Adaptive method. It particularly possesses three advantages over the 
AC-Adaptive method: \\
First and in contrast to RING, the AC-Adaptive method needs (nearly) 
opposed spokes to determine the 
shifts via correlation analysis. However, this requirement cannot be fulfilled 
for 
half circle acquisitions, 
where the projection angle is distributed in the 
range $\theta\in[0,\pi]$ \cite{Haji-Valizadeh_Magn.Reson.Med._2018}. We 
demonstrated this fact in 
numerical simulations (Fig. \ref{Fig:Sim}) for which the AC-Adaptive method 
provides inaccurate estimates for the gradient delays,
whereas RING yields accurate results.\\
Second, the 
need for opposed spokes in the AC-Adaptive method also prohibits the use of 
few spokes for gradient delay estimation without calibration scans, if at the 
same time a uniform k-space 
coverage shall be guaranteed. We found this notion in all experiments Fig.\ 
\ref{Fig:Sim}, \ref{Fig:Lego}, \ref{Fig:Invivo} which show pronounced streaking 
artifacts and gradient delay errors for few spokes using the AC-Adaptive 
method, 
while RING provides high quality results for any number of spokes.\\
Third, the AC-Adaptive method assumes that gradient delays solely 
translate 
into trajectory shifts in readout direction. This, however, is only the case 
for isotropic delays as shown in Fig.\ \ref{Fig:Trajplot}. In all other cases, 
the trajectory additionally experiences an orthogonal shift, which in 
particular means that even perfectly opposed spokes do no not cover the same 
k-space samples. This also explains why in Fig.\ \ref{Fig:Sim}a) the gradient 
delay error for the AC-Adaptive method only converges to zero in the isotropic 
case, 
but not in the oblique and axial case. \\
Recently, we have developed an extension to the AC-Adaptive method 
which 
allows gradient delay estimation from few spokes by exploiting the conjugate 
symmetry in k-space instead of finding opposed spokes \cite{Rosenzweig__2018}, 
which, however, suffers from the same model inconsistency concerning orthogonal 
shifts.\\
RING requires two parameters: $N_\text{pad}$, which determines the 
amount of k-space sub-sampling to increase accuracy and $\beta$ which defines 
the region in which the samples of the crossing spokes are compared to find the 
intersection 
point. In preliminary investigations (not shown) we found that the accuracy of 
the estimates does not significantly improve for $N_\text{pad}>100$, so we 
suggest and used $N_\text{pad}=100$ in all our experiments. In general, 
gradient delay induced k-space shifts are $<0.5\; 1/\text{FOV}$, thus the 
proposed value for $\beta=1.5$ is sufficient to find all intersection 
points. Note, that values 
$\beta<1$ may result in estimation inaccuracies as spokes that are not quite 
orthogonal could intersect outside of the so defined region. On the other hand, 
we 
recommend to never chose $\beta\gtrapprox4$, since samples outside of the 
central k-space contain less energy and are more affected by noise, which might 
suggest a false intersection point. In the scope of this work we only 
considered the intersection of a spoke with its most orthogonal counterpart, 
which is apparently enough to yield valid results. This restriction, however, 
can be 
relaxed and the intersection of a spoke with multiple other spokes can be 
considered 
in the fit Eq.\ (\ref{Eq:InverseProblem}), which can provide minor 
improvements for very few considered spokes.\\
RING proofs potential as a general, lightweight on- and offline gradient delay 
correction tool for radial imaging. Furthermore, it can be used for AC-Adaptive 
frame-by-frame gradient delay correction in interactive real-time MRI. Because 
of its flexibility, it can be directly applied to other k-space acquisition 
schemes based on radials, such as radial simultaneous multi-slice 
\cite{Wang_Magn.Reson.Imaging_2016,Rosenzweig__2017,Rosenzweig_Magn.Reson.Med._2018}
 or stack-of-stars 
\cite{Zhou_Magn.Reson.Med._2017,Block_J.KoreanSoc.Magn.Reson.Med._2014}.

\section{Conclusion}
We have presented a simple and straight forward new method dubbed RING to 
estimate gradient delay errors of radial trajectories from very few spokes. 
RING uses the gradient delay ellipse model introduced by Peters et 
al. 
\cite{Peters_Magn.Reson.Med._2003} and Moussavi et al.\
\cite{Moussavi_Magn.Reson.Med._2013} to fit the gradient delays Eq.\ 
(\ref{Eq:S}) using the intersection points of spokes. The method yields highly 
accurate and robust gradient delay estimates even for $N_\text{sp}=3$ spokes 
in vivo. For its data driven, auto-calibrating nature, it can simply be 
inserted 
as a module in existing online or offline frameworks without the need to adapt 
the measurement protocols. 

\section{Acknowledgements}
Supported by the DZHK (German Centre for Cardiovascular Research).  Part of 
this research was funded
by the Physics-to-Medicine Initiative Göttingen (LM der Niedersächsischen 
Vorab) and DFG (UE 189/1-1).

\appendix
\b{\section{Derivation of the gradient delay ellipse model}}\R{Ref 1.2-3}
First, we recall the definitions and results of 
\cite{Peters_Magn.Reson.Med._2003}. The gradients of the logical system
\begin{equation}
\bm{G}^\text{log}_\theta(t):= 
\left( 
\begin{array}{l} 
G^\text{log}_\text{read}(t) \cos\theta \\ 
G^\text{log}_\text{read}(t) \sin\theta \\ 
G^\text{log}_\text{slice}(t)
\end{array}
 \right),
 \label{Eq:GlogVector}
\end{equation}
can be transformed into the physical system
using the orthogonal transform
\begin{equation}
\bm{R}:=
\left(
\begin{array}{ccc}
R_{11} & R_{12} & R_{13} \\
R_{21} & R_{22} & R_{23} \\
R_{31} & R_{32} & R_{33} 
\end{array}
\right),
\end{equation}
which yields
\begin{equation}
G^\text{phy} (t)=\bm{R}G^\text{log}_\theta (t).
\label{Eq:PhyTrans}
\end{equation}
The timing delays $t_x$, $t_y$ 
and $t_z$, that effect the logical gradients, 
can be modeled using the delay 
operator
\begin{equation}
\mathcal{T}\bm{G}^\text{phy}(t) = 
\left(
\begin{array}{l}
G^\text{phy}_x(t-t_x)\\
G^\text{phy}_y(t-t_y)\\
G^\text{phy}_z(t-t_z)
\end{array}
\right).
\label{Eq:DelayOperator}
\end{equation}
Hence, the delayed gradients in the logical system can be obtained by
\begin{equation}
\tilde{\bm{G}}^\text{log}_\theta(t) = \bm{R}^T \mathcal{T} \bm{R} 
\bm{G}^\text{log}_\theta(t),
\end{equation}
where $^T$ denotes the transpose operation.
With definition 
\begin{equation}
\bm{T}:=\frac{\gamma}{2\pi} 
\left(
\begin{array}{ccc}
t_x & 0   & 0 \\
0   & t_y & 0 \\
0   & 0   & t_z
\end{array}
\right),
\end{equation}
the actual k-space shift is given by
\begin{align}
\delta \bm{k}_\theta
&=\frac{\gamma}{2\pi} \int_0^\tau 
(\tilde{\bm{G}}^\text{log}_\theta(t) - \bm{G}^\text{log}_\theta(t)) dt \\
&\overset{\text{Eq. (\ref{Eq:PhyTrans})}}{=}\frac{\gamma}{2\pi} \bm{R}^T 
\int_0^\tau 
(\mathcal{T}\bm{G}^\text{phy}(t) - 
\bm{G}^\text{phy}(t)) dt\\
&\overset{\text{Eq. \ref{Eq:DelayOperator}}}{=}\frac{\gamma}{2\pi} \bm{R}^T 
\int_0^\tau 
\left(
\begin{array}{c}
G_x^\text{phy}(t-t_x) - G_x^\text{phy}(t) \\
G_y^\text{phy}(t-t_y) - G_y^\text{phy}(t) \\
G_z^\text{phy}(t-t_z) - G_z^\text{phy}(t) \\
\end{array}
\right)dt\\
&\overset{\text{\cite{Peters_Magn.Reson.Med._2003}}}{=} \bm{R}^T 
\bm{T}(\bm{G^\text{phy}}(\tau) - \bm{G^\text{phy}}(0)) 
\label{Eq:PetersStep}\\ 
&\overset{\text{Eq. (\ref{Eq:PhyTrans})}}{=} \bm{R}^T \bm{T} 
\bm{R}(\bm{G}^\text{log}_\theta(\tau) - 
\bm{G}^\text{log}_\theta(0))\\
&\overset{\text{Eq. (\ref{Eq:GlogVector})}}{=} \bm{R}^T \bm{T} \bm{R} 
\left(
\begin{array}{l}
G^\text{log}_\text{read}(\tau) \cos\theta \\ 
G^\text{log}_\text{read}(\tau) \sin\theta \\ 
-G^\text{log}_\text{slice}(0)
\end{array}
\right).
\label{Eq:Last}
\end{align}
For Eq.\ (\ref{Eq:PetersStep}), 
we assumed that the temporal delays are small compared to the flattop time of 
the gradients. In Eq.\ (\ref{Eq:Last}), we used the fact that at the temporal 
center 
of the RF pulse ($t=0$), 
only the slice selection gradient is active and at the center of readout 
($t=\tau$) only the readout gradients are active. We refer the reader to 
\cite{Peters_Magn.Reson.Med._2003} for 
more details.

In general, we only have information about the projection direction Eq. 
(\ref{Eq:ProjDir}) 
$\hat{\bm{n}}^\text{log}_\theta=(\cos\theta, \sin\theta,0)^T$ and not 
about the actual 
gradient strength of a measurement. Therefore, we separate the projection 
direction using
\begin{equation}
\underline{\bm{G}}^\text{log} :=
\left(
\begin{array}{ccc}
G^\text{log}_\text{read}(\tau) & 0 & 0\\
0 & G^\text{log}_\text{read}(\tau) & 0\\
0 & 0 & -G^\text{log}_\text{slice}(0)
\end{array}
\right),
\end{equation}
and obtain
\begin{equation}
\delta \bm{k}_\theta = \bm{R}^T 
\bm{T} \bm{R} \underline{\bm{G}}^\text{log}(\hat{\bm{e}}^\text{log}_z + 
\hat{\bm{n}}^\text{log}_\theta),
\label{Eq:FullDeltak}
\end{equation}
with
$\hat{\bm{e}}^\text{log}_z=(0,0,1)^T$.

The first term of the right-hand-side of Eq.\ (\ref{Eq:FullDeltak}) is 
independent of the angle and thus does not effect the ellipse fit of RING and 
corresponds to a constant k-space offset, i.e. a 
linear phase in image space, and can therefore be neglected. Then, Eq.\ 
(\ref{Eq:FullDeltak}) written out is 
given by
\begin{gather}
\delta\bm{k}_\theta \approx \bm{R}^T\bm{T}\bm{R} 
\underline{\bm{G}}^\text{log}\hat{\bm{n}}^\text{log}_\theta \\
=
\begin{scriptsize}
\left(
\begin{array}{lll}
G^\text{log}_\text{read}(t_x R_{11}^2 + t_y R_{21}^2 + t_z R_{31}^2)
& G^\text{log}_\text{read}(t_x R_{11} R_{12} + t_y R_{21} 
R_{22} + t_z R_{31} R_{32}) 	
& G^\text{log}_\text{slice}(t_x R_{11} R_{13} + t_y R_{21} R_{23} + t_z 
R_{31} R_{33}) \\
G^\text{log}_\text{read} (t_x R_{12} R_{11} + t_y R_{22} R_{21} + t_z R_{32} 
R_{31}) 	
& G^\text{log}_\text{read}(t_x R_{12}^2 + t_y R_{22}^2 + t_z R_{32}^2)
& G^\text{log}_\text{slice} (t_x R_{12} R_{13} + t_y R_{22} R_{23} + t_z R_{32} 
R_{33}) \\
G^\text{log}_\text{read} (t_x R_{13} R_{11} + t_y R_{23} R_{21} + t_z 
R_{33}R_{31})
& G^\text{log}_\text{read} (t_x R_{12} R_{13} + t_y R_{22} R_{23} + t_z R_{33} 
R_{32})
& G^\text{log}_\text{slice} (t_x R_{13}^2 + t_y R_{23}^2 + t_z 
R_{33}^2)
\end{array}
\right)
\left(
\begin{array}{c}
\cos\theta\\
\sin\theta\\
0
\end{array}
\right)
\end{scriptsize}.
\end{gather}
Here, we are only interested in the in-plane gradient delays, for which it 
suffices to consider the top left $2\times2$ submatrix. By substitution we 
yield Eq.\ (\ref{Eq:deltak})
\begin{align}
\delta\bm{k}_\theta &\approx
\begin{scriptsize}
\left(
\begin{array}{lll}
G^\text{log}_\text{read}(t_x R_{11}^2 + t_y R_{21}^2 + t_z R_{31}^2)
& G^\text{log}_\text{read}(t_x R_{11} R_{12} + t_y R_{21} 
R_{22} + t_z R_{31} R_{32}) \\
G^\text{log}_\text{read} (t_x R_{12} R_{11} + t_y R_{22} R_{21} + t_z R_{32} 
R_{31}) 	
& G^\text{log}_\text{read}(t_x R_{12}^2 + t_y R_{22}^2 + t_z R_{32}^2)
\end{array}
\right) \hat{\bm{n}}^\text{log}_\theta 
\end{scriptsize}\\
&:=
\left( 
\begin{array}{cc}
S_x & S_{xy} \\
S_{xy} & S_y 
\end{array}
\right)\hat{\bm{n}}^\text{log}_\theta.
\end{align}

\printendnotes


\bibliography{radiology}

\iftoggle{MRM}{}
{


\section*{Supplementary material}

# 1 | ACCURACY OF THE INTERSECTION POINT DETERMINATION

We perform phantom simulations with different numerical phantoms, signal to noise ratios and numbers of coils to analyze the accuracy of the intersection point determination using a pixel-wise comparison of the spokes' samples.

We utilize a Shepp-Logan phantom and two other geometric phantoms. The left side of Sup. Fig. 1 shows NUFFT reconstructions of the phantoms for different signal to noise ratios. We simulate the corresponding delayed radial k-space data ($s := (0.3, -0.1, 0.2)$, full circle golden angle trajectory, 160 read-out samples, 159 spokes) for $c = 1, \ldots, 8$ coils and add Gaussian white noise. As a measure for the noise we divide the energy of the noise-free k-space by the energy of the added noise: $\text{SNR}_k = E_s/E_n$). We compare all actual intersection points with the estimated intersection point using Eq. (4,5) and $s$. For the estimation, the values for $a_{\theta_i}$ are determined by pixelwise comparison of the spokes' samples as described in the manuscript. The analytical expression used to determine the actual intersection point can be derived using Eq. (5):

$$a_{\theta_i} = \left( \frac{\hat{n}_{\theta_j,2}}{\hat{n}_{\theta_j,2} \hat{n}_{\theta_j,1}} - \hat{n}_{\theta_j,1} \right) \left( S_x \xi_1 + S_{xy} \xi_2 - \frac{\hat{n}_{\theta_j,1}}{\hat{n}_{\theta_j,2}} (S_{xy} \xi_1 + S_y \xi_2) \right).$$

The central column of Sup. Fig. 1 shows the root-mean-square (RMS) error between all measured and actual intersection points for different noise levels and numbers of coils.

Furthermore, we estimate the gradient delays $s^{\text{est}}_{\text{SNR}_k,c}$ for the same numbers of coils $c$ and signal to noise ratios. The gradient delay error $\mathcal{E}(s, s^{\text{est}}_{\text{SNR}_k,c})$ (Eq. (12)) is depicted in Sup. Fig. 1, right column.

Both the gradient delay error and the RMS error of the intersection points improve with the number of utilized coils. This behavior is sensible as samples from all channels can be used to determine the intersection points and thus more robust results can be obtained. While a single coil yields bad results for any noise level, the use of multiple coils, which today is clinical standard, yield accurate results even for low $\text{SNR}_k$ values. The accuracy is comparable for all investigated phantoms.

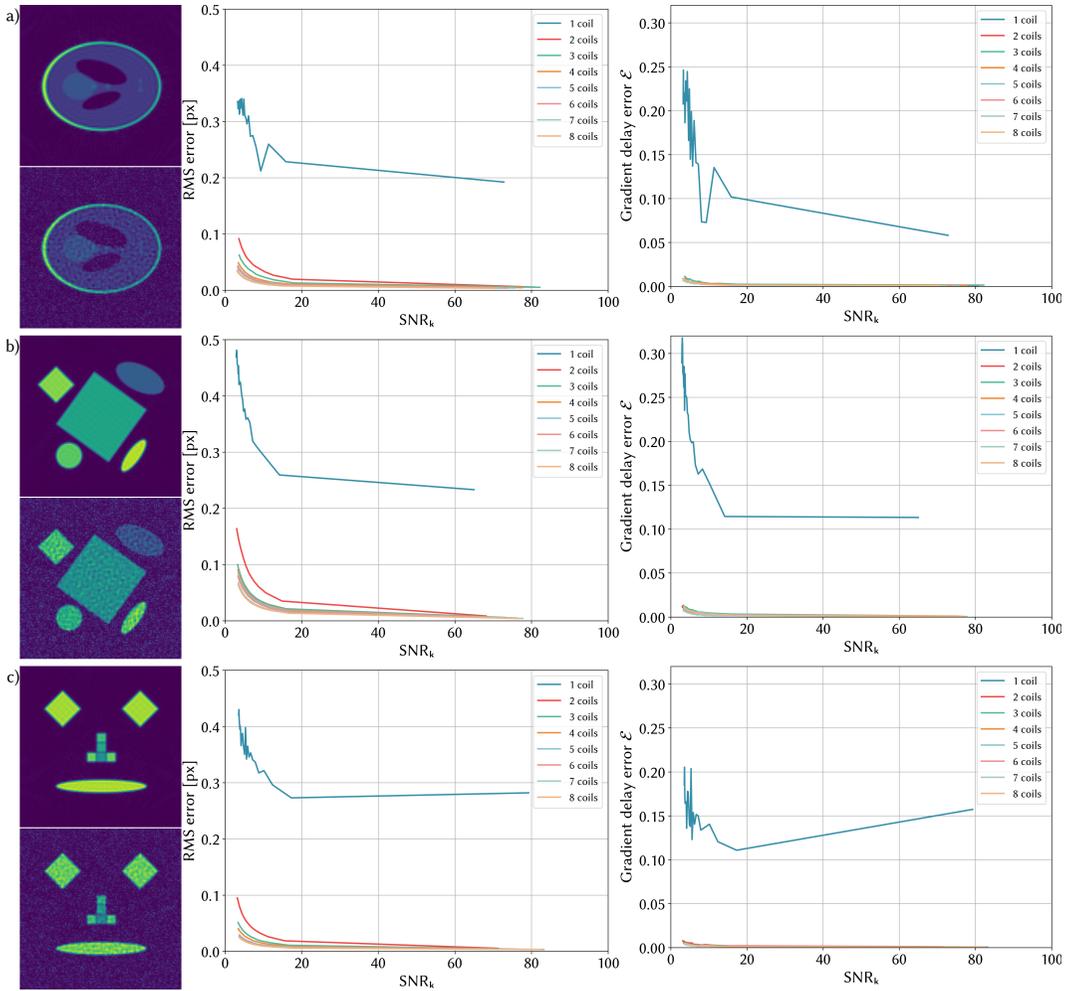

**SUP. FIGURE 1** Intersection point accuracy analysis for three different numerical phantoms a), b), c). Left: NUFFT reconstructions without noise and with $SNR_k \approx 7$. Center: Root-mean-square error of the measured intersection point against the $SNR_k$ for different numbers of coils. Right: Gradient delay estimation error against the $SNR_k$ for different numbers of coils.


}

\end{document}